\documentclass{nature}

\usepackage{graphicx}
\usepackage{dcolumn}
\usepackage{bm}
\usepackage{multirow}
\usepackage{lineno}
\usepackage{amsmath}

\makeatletter
\let\saved@includegraphics\includegraphics
\AtBeginDocument{\let\includegraphics\saved@includegraphics}
\renewenvironment*{figure}{\@float{figure}}{\end@float}
\makeatother

\title{Magnetic Skyrmion States in Janus Monolayers \\ of Chromium Trihalides Cr(I,X)$_3$ }
\author{Changsong Xu$^{1}$, ~Junsheng Feng$^{2}$, ~Hongjun Xiang$^{3,4,*}$ ~\& ~L. Bellaiche$^{1,*}$}

\begin{document}

\maketitle

\begin{affiliations}
 \item Physics Department and Institute for Nanoscience and Engineering, University of Arkansas, Fayetteville, Arkansas 72701, USA
 \item School of Physics and Materials Engineering, Hefei Normal University, Hefei 230601, P. R. China
 \item Key Laboratory of Computational Physical Sciences (Ministry of Education), State Key Laboratory of Surface Physics, and Department of Physics, Fudan University, Shanghai, 200433, China
 \item Collaborative Innovation Center of Advanced Microstructures, Nanjing 210093, P. R. China
\end{affiliations}

\begin{abstract}
  Magnetic skyrmions are nano-scale spin structures that are promising for  ultra-dense memory and logic devices. Recent progresses in two-dimensional magnets encourage the idea to realize skyrmionic states in freestanding monolayers. However, monolayers such as  CrI$_3$ lack  Dzyaloshinskii-Moriya interactions (DMI) and thus do not naturally exhibit skyrmions but rather a ferromagnetic state.
  Here we propose the fabrication of Cr(I,X)$_3$ Janus monolayers, in which the Cr atoms are covalently bonded to the underlying I ions and top-layer Br or Cl atoms. By performing first-principles calculations and Monte-Carlo simulations, we identify strong enough DMI, which leads to not only helical cycloid phases, but also to intrinsic skyrmionic states in Cr(I,Br)$_3$ and magnetic-field-induced skyrmions in Cr(I,Cl)$_3$.
\end{abstract}

\newpage

Magnetic skyrmions are nano-scale spin clusters with topological stability, and are promising for advanced spintronics \cite{nagaosa2013topological,fert2013skyrmions}. One requirement toward such applications is that the hosting materials should be thin films, so that the nano size of skyrmions can be taken full advantage of. Besides previous studies on bulk MnSi \cite{heinze2011spontaneous,muhlbauer2009skyrmion,neubauer2009topological,pappas2009chiral},
recent works focused on ultrathin films, such as FeGe \cite{yu2011near,huang2012extended} and rare-earth ion garnet \cite{avci2019interface,shao2019topological}, which both take advantage of the  Dzyaloshinskii-Moriya interaction (DMI) arising from the heavy metal substrate. However, no skyrmionic state has ever been reported to intrinsically exist in free-standing {\it monolayers}, to the best of our knowledge, while two-dimensional (2D) semiconducting magnets, such as monolayer CrI$_3$ \cite{huang2017layer}, are recently attracting much attention due to their novel physics and rich applications \cite{gibertini2019magnetic}.
The ferromagnetic monolayer CrI$_3$ crystalizes in honeycomb lattice made of edge-sharing octahedra. Its ferromagnetic order is stabilized by an out-of-plane anisotropy \cite{huang2017layer}, which arises from single ion anisotropy (SIA) and Kitaev-type exchange coupling that both result from the SOC of its heavy ligands \cite{xu2018interplay,lado2017origin}. However, the ingredient DMI is absent between the most strongly coupled first nearest neighbor (1st NN) Cr-Cr pairs, because the inversion center between the two Cr atoms prevents its existence \cite{moriya1960anisotropic}. Interestingly,  very recent theoretical study proposed the application of electric field to break the inversion center and induce DMI in monolayer CrI$_3$  \cite{behera2019magnetic}. Although this clever method leads to CrI$_3$ monolayers becoming closer to adopt a skyrmion phase, the weak effects of electric field in generating DMI, as well as the rather strong out-of-plane anisotropy, hinders the actual creation of skyrmions in this system. 

Here we propose a more effective approach that consists in fabricating Janus monolayers of chromium trihalides Cr(I,X)$_3$ (X = Br, Cl). One example of Janus monolayers is the transition metal dichalcogenides MoSSe, which originates from the well-known monolayer MoS$_2$ but with one layer of S being substituted by Se.
Such Janus monolayer is experimentally achievable, since MoSSe has been reproducibly obtained using different methods \cite{lu2017janus,zhang2017janus}, which strongly suggests the feasibility of also creating Cr(I,X)$_3$ Janus monolayers.
In this manuscript, we apply density functional theory (DFT) and parallel tempering Monte Carlo (PTMC) simulations to study the magnetic interactions and to investigate skyrmionic states of two Janus monolayers, namely Cr(I,Br)$_3$ and Cr(I,Cl)$_3$, as well as the prototype CrI$_3$ (see structural schematics in Figs. 1a and 1b). As we will show, these Janus monolayers exhibit not only strong enough DMI, but also a decrease in magnetic anisotropy energy (MAE), which both contribute to stabilizing skyrmionic states. We will also demonstrate that the presence of Kitaev interaction and the application of magnetic field both benefit to the emergence of skyrmions in these Cr(I,X)$_3$ Janus monolayers.


\noindent
{\bf Results}

\noindent
{\bf Hamiltonian and magnetic parameters}

We consider the following Hamiltonian for describing the magnetic interactions of the Cr(I,Br)$_3$ and Cr(I,Cl)$_3$ Janus monolayers, as well as the prototype CrI$_3$ monolayer, using the generalized matrix form up to the second nearest neighbors (2nd NN) for the spins, $\bm{{\rm S}}$:
\begin{equation}\label{Eq1}
  \mathcal{H} = \frac{1}{2} \sum_{n=1,2}~ \sum_{(i,j)_n} \bm{{\rm S}}_i {\cdot} \mathcal{J}_{n,ij} {\cdot} \bm{{\rm S}}_j
  +\sum_{i} \bm{{\rm S}}_i {\cdot} \mathcal{A}_{ii} {\cdot} \bm{{\rm S}}_i ~,
\end{equation}
where the first term is the exchange coupling that runs over all first $(i,j)_1$ and second $(i,j)_2$ NN Cr pairs, respectively, and the second term represents SIA that runs
over all Cr sites. $\mathcal{J}$ and $\mathcal{A}$ are 3$\times$3 matrices, for which the elements can be extracted using the four-state energy mapping method and density-functional theory (DFT) (see Method section and Ref.  \cite{xu2018interplay,xu2019magnetic,xiang2011predicting,xiang2013magnetic} for details).
The $\mathcal{J}$ matrices can be further decomposed into an isotropic parameter $J'$ (to be defined later), the symmetric Kitaev term $K$ \cite{xu2018interplay} and the antisymmetric DMI vector $\bm{{\rm D}}$ \cite{xu2019magnetic}; and the $\mathcal{A}$ matrix simply reduces to  its $A_{zz}$ components by symmetry. The Hamiltonian can then be rewritten as
\begin{equation}\label{Eq2}
  \mathcal{H} = \frac{1}{2} \sum_{n=1,2}~ \sum_{(i,j)_n}
  \{J'_n \bm{{\rm S}}_i {\cdot}  \bm{{\rm S}}_j
  + K_n S_i^{\gamma} S_j^{\gamma}
  + \bm{{\rm D}}_{n,ij} {\cdot} (\bm{{\rm S}}_i {\times}  \bm{{\rm S}}_j)\}
  + \sum_{i} A_{zz} (S_i^{z})^2 ~.
\end{equation}
where $S^{\gamma}$ is the $\gamma$ component of \bm{{\rm S}}
in a local basis \{$\alpha,\beta,\gamma$\} that is related to the Kitaev interaction. Such local basis, as well as the direction of DMI vector, are  schematized in Figs. 1c and 1d, and will be discussed in detail in the following paragraphs.

Note that the Kitaev interaction is a bond-wised anisotropic exchange coupling that was first proposed by A. Kitaev \cite{kitaev2006anyons} and then found in Na$_2$IrO$_3$ by G. Jackeli and G. Khaliullin \cite{jackeli2009mott}.
A non-negligible Kitaev coefficient has also been previously identified in CrI$_3$ and CrGeTe$_3$ and was found to be crucial for determining and understanding the different magnetic anisotropies of these two latter compounds \cite{xu2018interplay}. A Kitaev term was also needed in order to reproduce the temperature evolution of magnetization in CrBr$_3$ few layers \cite{kim2019hall} and is the key ingredient to create  quantum spin liquid states (when the $J'$ parameter is vanishing \cite{chaloupka2010kitaev}).

In the present case, for the 1st NN of Cr(I,X)$_3$, as well as that of CrI$_3$, diagonalization of the symmetric part of $\mathcal{J}_1$ matrix leads to (i) nearly degenerate lower-in-energy $J'_{1,\alpha}$ and $J'_{1,\beta}$, where the eigenvectors $\alpha_1$ and $\beta_1$ lie in the Cr$_2$L$_2$ (L for ligand) plane; and (ii) higher-in-energy $J'_{1,\gamma}$  being associated with the perpendicular $\gamma_1$ axis (see Figs. 1c and 1d).
Similarly for the 2nd NN, the low-energy easy plane is spanned by the perpendicular 3rd NN Cr pair ($\alpha_2$ axis) and the two most related ligands ($\beta_2$ axis); the hard axis ($\gamma_2$ axis) is thus perpendicular to the easy plane (see schematics in Figs. 1c and 1d).
For both the 1st and the 2nd NN, the aforementioned isotropic exchange parameter is thus defined as $J'=(J'_{\alpha}+J'_{\beta})/2$ and the Kitaev coefficient as $K=J'_{\gamma}-J'$. Note that we further define another isotropic parameter as $J=(J_{xx}+J_{yy}+J_{zz})/3$, which is equivalent to the $J$ coefficient involved in the commonly used $J\bm{{\rm S}}_i {\cdot}  \bm{{\rm S}}_j$ term. Such isotropic $J$ is thus different from $J'$ and its ratio with the DMI parameter ($|D/J|$) can provide information about the existence of magnetic skyrmions or not.

Furthermore, the directions of the DMI vectors are also distinct between 1st and 2nd NN. Specifically and as further illustrated in Figs. 1c and 1d, the $\bm{{\rm D}}_1$ vectors of all Cr(I,X)$_3$ are basically parallel to $\gamma_1$, which satisfy Moriya's rule \cite{moriya1960anisotropic} and thus testify the accuracy of our calculations. On the other hand, the $\bm{{\rm D}}_2$ vectors all stay in the ($\alpha_2,\beta_2$) easy plane and are close to the $\beta_2$ axis.

Let us now look at the precise values of the magnetic parameters obtained from DFT and four-state method \cite{xu2018interplay,xu2019magnetic,xiang2011predicting,xiang2013magnetic}.
As shown in Table I, both $J_1$ and $J_2$ yield negative values, which imply ferromagnetism (FM) for all investigated systems, which is consistent with the measured ferromagnetism of CrX$_3$, with  X = Cl, Br and I \cite{morosin1964x,kim2019hall,huang2017layer}. The $J_1$, $K_1$ and $A_{zz}$ parameters all decrease in magnitude when the X ion of Cr(I,X)$_3$ varies from I to Cl, via Br. This decrease in $J_1$ results from the shrinking of lattice constants (leading to shortened Cr-Cr distance and enhanced direct antiferromagnetic exchange), while that of $K_1$ and $A_{zz}$ root in the weakening of the SOC strength (as consistent with the location of I, Br and Cl in the periodic Table). On the other hand, the $J_2$ and $K_2$ coefficients show no significant changes with the X ion.
Moreover, the Janus monolayers Cr(I,Br)$_3$ and Cr(I,Cl)$_3$ exhibit remarkable $D_1$ values of 0.270 meV and 0.194 meV, respectively, while the prototype CrI$_3$ has no finite $D_1$, as aforementioned. Such values, altogether with $J_1$, yield large $|D_1/J_1|$ ratios of 0.150 for Cr(I,Br)$_3$ and 0.194 for Cr(I,Cl)$_3$, which are within the typical range of 0.1-0.2 known to generate skyrmionic phases \cite{fert2013skyrmions}. Such $|D_1/J_1|$ ratios are much larger than the reported value 0.071, obtained when applying an extremely large 2 V/nm electric field to CrI$_3$ monolayer \cite{behera2019magnetic}. Notably, our three investigated systems also show non-negligible $D_2$ values, which leads to $|D_2/J_2|$ ratios comparable to their 1st NN counterparts, as detailed in Table I. The significant $|D_1/J_1|$ and $|D_2/J_2|$ ratios of Cr(I,Br)$_3$ and Cr(I,Cl)$_3$ demonstrate that the fabrication of Janus monolayers is an effective way to induce considerable DMI and is thus promising to create magnetic skyrmions.

Another important factor that affects the formation of skyrmions is the magnetic anisotropy energy (MAE), which is defined as the energy difference between out-of-plane FM (zFM) and in-plane FM (xFM) states, $\epsilon=E_{zFM}-E_{xFM}$, where $\epsilon < 0$ ($> 0$, respectively) favors out-of-plane (in-plane, respectively) FM. Here, for Cr(I,X)$_3$ and CrI$_3$, the total anisotropy is the result of the interplay between Kitaev interaction and SIA \cite{xu2018interplay} and is calculated using the model energy of Eq. (2) with the parameters of Table 1.
As shown in Fig. 2a, CrI$_3$, which has an $\epsilon$ = -0.693 meV/Cr, favors strong out-of-plane anisotropy and Cr(I,Br)$_3$, with $\epsilon$ = -0.218 meV/Cr, exhibits mild out-of-plane anisotropy. In contrast, Cr(I,Cl)$_3$ shows weak in-plane anisotropy since $\epsilon$ = 0.111 meV/Cr (note that Fig. 2a also displays the $\epsilon$ directly calculated from DFT, that agree rather well with those obtained by the model -- which thus further testifies the accuracy of our model Hamiltonian). As we will see latter, such differences in anisotropy between Cr(I,Br)$_3$ and Cr(I,Cl)$_3$  plays a role on the morphology of the skyrmions.

\noindent
{\bf Intrinsic skyrmionic states in Cr(I,Br)$_3$ Janus monolayer}

As detailed in the Method section, parallel tempering Monte Carlo (PTMC) simulations using the Hamiltonian of Eq. (1)  are performed over a 50$\times$50$\times$1 supercell to find spin structures with low energies. These latter spin structures are then further relaxed with the conjugate gradient (CG) method, until arriving at energy minima. Such optimization scheme guarantees the converged spin structures to be either metastable states or the ground state at the temperature of zero Kelvin.

Practically, the ground state of Cr(I,Br)$_3$ is determined to be out-of-plane ferromagnetism (zFM), of which the energy is set to be zero.
The first metastable state (the convention of terminology here is that the $n$th metastable state has the $(n+1)$th lowest energy) of Cr(I,Br)$_3$ is an out-of-plane cycloid structure, as shown in Fig. 3a, with an energy of 0.056 meV/Cr. Its propagation direction is symmetrically equivalent along $\bm{{\rm (a+2b)}}$, $\bm{{\rm (2a+b)}}$ and $\bm{{\rm (b-a)}}$. Such cycloid structure yields no finite topological charge $Q$ that is defined as
\begin{equation}\label{Eq3}
Q=\frac{1}{4\pi }\int \bm{{\rm m}} \cdot (\frac{\partial \bm{{\rm m}}}{\partial x})\times (\frac{\partial \bm{{\rm m}}}{\partial y})dxdy~,
\end{equation}
where $\bm{{\rm m}}$ is the unit vector lying along the local magnetic moment's direction.

On the other hand, skyrmionic states with finite $Q$ are found as the second and even higher metastable states. As shown in Fig. 3b, anomalous spin patterns occur at the boundary of dark and bright zones, which are actually the domains with spins having ``in'' and ``out'' {\it out-of-plane} components , respectively, i.e., z-components that are negative and positive, respectively, along the z-axis (note that, on the other hand, the arrows displayed in Fig.3 represent the {\it in-plane} components of the magnetic dipoles). We name such boundary as domain wall.  The novel spin structure of Fig. 3b renders $Q=1$ and is thus topologically identical to the common bubble-like skyrmion \cite{heinze2011spontaneous,muhlbauer2009skyrmion,neubauer2009topological,pappas2009chiral}.
Such skyrmionic state has an energy of 0.069 meV/Cr (with respect to the ground state) and is also 0.013 meV/Cr higher than the cycloid state in our used 50$\times$50$\times$1 supercell. Moreover, it is found that either the upper or the lower domain wall can host multiple skyrmions. For example, state possessing two skyrmions ($Q=2$) in the supercell can have (i) one skyrmion at each domain wall (Fig. 3c) or (ii) two skyrmions at the same domain wall (Fig. 3d).
Interestingly, such two cases possess degenerate energy of 0.082 meV/Cr.
In fact, the energy is  a linearly increasing function with simply the number of skyrmions, as shown in Fig. 2c, with the maximum number of skyrmions existing within the 50$\times$50$\times$1 supercell being four (see Fig. 3e for that four-skyrmion state).

Furthermore, it is found that the domain wall can host not only skyrmions, but also anti-skyrmions. As a matter of fact and as shown in Fig. 3f, (i) the spins rotate {\it clockwise} from {\it left to right} along the {\it lower} domain wall, which renders $Q=2$ and characterizes two skyrmions, as similar to the case of Fig. 3d; (ii) but, in contrast, the spins rotate {\it anticlockwise} from {\it right to left} along the {\it upper} domain wall, which yields $Q=-2$ and indicates two antiskyrmions.
The total topological charge of Fig. 3f is thus $Q=0$. Comparing Fig. 3e and 3f, we find that either domain wall (upper or lower) can host either skyrmions or anti-skyrmions.
The energies of the two spin configurations in Figs. 3e and 3f are further found to be exactly the same, indicating that skyrmion and anti-skyrmion are energetically degenerate, which has never been reported, to the best of our knowledge. Besides, the existence of antisyrmions is actually rather rare in surface or interface systems \cite{hoffmann2017antiskyrmions}. It is also found here that skyrmions and anti-skyrmions can not exist at the same domain wall, since they tend to annihilate with each other which would in fact lead to the transformation to the cycloidal state (resulting in $Q=0$).


\noindent
{\bf Magnetic field induced skyrmion states in Cr(I,Cl)$_3$ Janus monolayer.}

Let us now turn our attention to the Cr(I,Cl)$_3$ system.
As consistent with its positive MAE, the ground state of Cr(I,Cl)$_3$ is determined to be in-plane zigzag-canted FM, as shown in Fig. 4a. The spin vectors at the A and B sites of the honeycomb lattice make an angle of 6.7$^\circ$.  Such spin canting is further evidenced by DFT calculations on a unit cell, as a moment of 0.3 $\mu_B$ along the $y$-direction emerges from the initial FM state fully lying along the x-direction (xFM), with an energy lowering of 0.024 meV/Cr with respect to that xFM state (note that the resulting angle between spins is equal to 6.2$^\circ$ in DFT calculations, which compares rather well with the aforementioned 6.7$^\circ$ given by numerical simulations using the magnetic Hamiltonian of Eq. (2)). Such zigzag-canted FM state is degenerate when the main component of the magnetization is along the $\bm{{\rm a}}$ direction (which is precisely our x-direction), or along $\bm{{\rm b}}$ and $\bm{{\rm (a+b)}}$ directions.

The first metastable state is a mostly in-plane cycloidal structure superimposed with small out-of-plane components forming a spin wave, as shown in Fig. 4b. Note that such spin arrangement is reminiscent of the well-known complex spin organization of BiFeO$_3$ that consists in the coexistence of an in-plane cycloidal structure  and an out-of-plane spin density wave  \cite{rahmedov2012magnetic}.
Such cycloidal structure in Cr(I,Cl)$_3$ has an energy higher by 0.026 meV/Cr from the ground state and is degenerate for propagation directions being along $\bm{{\rm (a+2b)}}$, $\bm{{\rm (2a+b)}}$ or $\bm{{\rm (b-a)}}$ directions. Interestingly, skyrmionic states are not ``naturally'' stable in Cr(I,Cl)$_3$  within our Hamiltonian using the original DFT-derived magnetic parameters.
On the other hand, one can stabilize skyrmionic states there by applying an out-of-plane magnetic field $B$ in order to compensate the in-plane anisotropy. It is numerically found that a critical field of 0.65 T tunes the MAE from positive to negative. However, the magnetic field ($<$ 2 T) does not polarize the spins to the fully out-of-plane direction, but rather renders the in-plane zigzag-canted FM now possessing an out-of-plane component, as shown in Fig. 4c. The previously metastable cycloid state can not survive and transform to such FM state when $B >$ 0.2 T. Practically,  skyrmions are numerically found to be stabilized as metastable states when the magnetic field ranges between 0.5 T and 1.3 T. For instance and as shown in Fig. 4d for a magnetic field of 0.8 T, one skyrmion, with a very low energy of 0.015 meV/Cr, is created as a metastable state, rendering $Q = 1$. Such skyrmion is quite similar to the common bubble-like skyrmions \cite{heinze2011spontaneous,muhlbauer2009skyrmion,neubauer2009topological,pappas2009chiral}, but also possesses some unique characteristics. For instance,  it is not isotropic in the plane and there is always a bright (spin-out) zone being at one side of the dark (spin-in) skyrmion center, as shown in Figs. 4d-f.
In fact, up to three skymions can be stabilized within the 50$\times$50$\times$1 supercell and the energy also linearly  increases with the skymion numbers in Cr(I,Cl)$_3$, as shown in Fig. 2d. Interestingly, we further find that, wherever the skyrmions are initialized in the supercell, after optimization, their core tend to be aligned equivalently along $\bm{{\rm a}}$, $\bm{{\rm b}}$ and $\bm{{\rm (a+b)}}$ directions, which are the degenerate directions for the in-plane zigzag-canted FM state, as shown in Figs. 4e and 4f. Such well organized structure may be of benefit for potential application in memory devices.

\noindent
{\bf Discussion and Conclusion}

Note that we also studied the effects of Kitaev interactions in our investigated systems. It is found that, if the Kitaev terms are turned off in our considered Hamiltonian, the skyrmionic state can no longer be stabilized in Cr(I,Br)$_3$ Janus monolayers. As shown in Fig. 2b, $K_1$ (respectively, $K_2$) results in negative (respectively, positive) MAE and thus contributes to out-of-plane (respectively, in-plane) anisotropy. The interplay between Kitaev terms and SIA is responsible for the total anisotropy, which plays an important role in the morphology of skyrmions. For instance, we numerically found that arbitrarily modifying the MAE  gives the following results:  (i) a mild out-of-plane anisotropy benefits the formation of skyrmions, while (ii) a too small anisotropy results in skyrmions with large diameter, as shown in Fig. S1.
Moreover, the frustration arising from Kitaev interaction \cite{xu2018interplay} adds to the disorder of the system, which also facilitate the formation of skyrmions.
We also studied an analogous system, that is Cr(Br,Cl)$_3$. It exhibits in-plane zigzag-canted FM ground state and a metastable in-plane cycloid, as well as magnetic field induced skyrmion states, which makes it very much similar to Cr(I,Cl)$_3$.

In summary, we proposed the fabrication of Cr(I,X)$_3$ (X = Br, Cl) Janus monolayers to induce large DMI and subsequent magnetic skyrmion states. By combining DFT and MC simulations, we find that Cr(I,Br)$_3$ can intrinsically host metastable skyrmionic phases, while a skyrmionic state of Cr(I,Cl)$_3$ can be stabilized by applying an out-of-plane magnetic field. Our study thus suggests a feasible approach to create skyrmions in semiconducting magnets consisting of  chromium trihalides Janus monolayers. Such  presently predicted skyrmionic phases are not only useful for memory and logic devices, but can also be promising for energy storage using topological spin textures \cite{xiao2018energy}.


\clearpage
\begin{methods}
{\bf DFT calculations} are performed using the Vienna ab-initio simulation package (VASP) \cite{prb.59.1758}. The projector augmented wave (PAW) method \cite{blochl1994projector} is employed with the following electrons being treated as valence electrons:  Cr 3$p$, 4$s$ and 3$d$, I 5$s$ and 5$p$, Br 4$s$ and 4$p$, and Cl 3$s$ and 3$p$. The plane wave energy cut off is chosen to be  350 eV. The length of the $c$ axis of all investigated systems is set to be 19.807 \AA~, which yields thick enough vacuum layers of more than 16.5 \AA. The local density approximation (LDA) \cite{kohn1965self} is used, with an effective Hubbard $U$ parameter chosen to be 0.5 eV for the localized 3$d$ electrons of Cr ions \cite{xu2018interplay}. $k$-point meshes are chosen such as they are commensurate with the choice of 4$\times$4$\times$1 for the unit cell  (that contains 10 atoms). The choices on LDA, Hubbard $U$ and $k$-point meshes were previously tested to be accurate \cite{xu2018interplay}. The Hellman-Feynman forces are taken to be converged when they become smaller than 0.001 eV/\AA~on each ion.

\noindent
{\bf The four-state energy mapping method} is applied to obtain the elements of the $\mathcal{J}$ and $A_{zz}$ matrices  \cite{xiang2011predicting,xiang2013magnetic,xu2018interplay,xu2019magnetic}. The four-state method considers one specific magnetic pair (or site) at a time. Without loss of generality, the total energy can be written as:
\begin{linenomath*}
\begin{equation}
  E = E_{spin} + E_{nonspin} = E_0 + \bm{{\rm S}}_1{\cdot}\mathcal{J}_{12}{\cdot}\bm{{\rm S}}_2 + \bm{{\rm S}}_1 {\cdot} \bm{{\rm K}}_1
  + \bm{{\rm S}}_2 {\cdot} \bm{{\rm K}}_2
\end{equation}
\end{linenomath*}
where $\bm{{\rm S}}_1{\cdot}\mathcal{J}_{12}{\cdot}\bm{{\rm S}}_2$ is the exchange coupling between the magnetic site 1 and site 2; $\bm{{\rm S}}_1 {\cdot} \bm{{\rm K}}_1$ ($\bm{{\rm S}}_2 {\cdot} \bm{{\rm K}}_2$, respectively) represents the coupling between site 1 (site 2, respectively) and all the other magnetic sites that are different from sites 1 and 2; and the energy from interactions among all these other magnetic sites, together with non-magnetic energy $E_{nonspin}$, is gathered in $E_0$.
To calculate, e.g., the $J_{xy}$ component of $\mathcal{J}_{12}$, four spin states are considered for the \{Cr1,Cr2\} pair: \{$(S~0~0),(0~S~0)$\}, \{$(S~0~0),(0~-S~0)$\}, \{$(-S~0~0),(0~S~0)$\} and \{$(-S~0~0),(0~-S~0)$\}; while other spins are perpendicular to those of Cr1 and Cr2. The $J_{xy}$ can then be constructed with the energies obtained from DFT calculations using the following equation:
\begin{linenomath*}
\begin{equation}
  J_{xy} = \frac{E_1 - E_2 -E_3 +E_4}{4S^2}
\end{equation}
\end{linenomath*}

Such four-state method is accurate when using a large enough supercell (to prevent the coupling between some sites present in the supercell and periodically  repeated sites from happening). For example, SIA and the 1st NN parameters are calculated using a 2$\times$2$\times$1 supercell and the 2nd NN coefficients using a 3$\times$3$\times$1 supercell. See Ref. \cite{xiang2011predicting,xiang2013magnetic,xu2018interplay,xu2019magnetic} for more details.

\noindent
{\bf Parallel tempering Monte Carlo (PTMC) simulations and conjugate gradient (CG) method.}
PTMC simulations with heat bath algorithm \cite{miyatake1986implementation} are performed using the Hamiltonian of Eq. (1) to update the spin structures. The results shown in the manuscript are based on a 50$\times$50$\times$1 supercell, while a 54$\times$54$\times$1 supercell is also adopted, which leads to similar skyrmion states. 200,000 MC sweeps are performed at each temperature.
After the MC simulations, a CG method \cite{hestenes1952methods} is applied to further optimize the spin configuration. Specifically, the directions of spins are described with $(\theta_i,\phi_i)$, which are independent variables. Such $(\theta_i,\phi_i)$ can thus be optimized locally by the CG method to minimize the force on each spin.
The energy convergence criteria is set to be 1E-6 eV. The combination of PTMC simulations and the CG method guarantees that our reported magnetic configurations are all located at global/local energy minima.

\noindent
{\bf Calculation of topological charge $Q$.}
Although Eq. (3) is commonly used to define topological charge $Q$, it is not convenient to employ it to calculate $Q$ for discrete lattice of spins $\bm{n}_i$ $(n_{i,x},n_{i,y},n_{i,z})$. Consequently, we alternatively adopt the definition of Berg and L$\rm \ddot{u}$stcher \cite{berg1981definition},
\begin{equation}
  Q=\frac{1}{4\pi} \sum_l A_l
\end{equation}
\begin{equation}
  {\rm cos}(\frac{A_l}{2})=\frac{1 + \bm{n}_i\cdot\bm{n}_j + \bm{n}_j\cdot\bm{n}_k + \bm{n}_k\cdot\bm{n}_i}{\sqrt{2(1+\bm{n}_i\cdot\bm{n}_j)(1+\bm{n}_j\cdot\bm{n}_k)(1+\bm{n}_k\cdot\bm{n}_i)}}
\end{equation}
\begin{equation}
  {\rm sign}(A_l)={\rm sign}[\bm{n}_i\cdot(\bm{n}_j\times\bm{n}_k)]
\end{equation}
where $l$ runs over all elementary triangles made of neighboring spin sites and $A_l$ is the solid angle formed by the three spin vectors $\bm{n}_i$, $\bm{n}_j$ and $\bm{n}_k$ of the $l$th triangle. Note that the triangles should cover all lattice area with no overlap and $\bm{n}_i$, $\bm{n}_j$ and $\bm{n}_k$ should be anticlockwisely ordered. See details for exceptional cases in Refs. \cite{berg1981definition,muller2019spirit}.


\end{methods}

\begin{addendum}
 \item[Data Availability] All data generated or analysed during this study are included in this published article (and its supplementary information files).
 \item[Acknowledgements]This work is supported by the Office of Basic Energy Sciences under contract ER-46612. H.X. is supported by NSFC (11374056), the Special Funds for Major State Basic Research (2015CB921700), Program for Professor of Special Appointment (Eastern Scholar), Qing Nian Ba Jian Program, and Fok Ying Tung Education Foundation. J.F. acknowledges the support from Anhui Provincial Natural Science Foundation (1908085MA10). The Arkansas High Performance Computing Center (AHPCC) is also acknowledged.
 \item[Author contributions] C.X. and J.F. contribute equally to this work. The idea of this paper is conceived during the discussion of all authors. C.X. performed the calculations. H.X. and L.B. supervised this work. C.X. wrote the first draft of the manuscript.  All authors contributed to the discussion of the results and comments on the first draft of the manuscript.
 \item[Additional Information] See supplementary information for details about our predictions.
 \item[Competing Interests] The authors declare that they have no competing financial interests.
 \item[Correspondence] Correspondence should be addressed to
        H.X. (email: hxiang@fudan.edu.cn) or L.B. (email: laurent@uark.edu).
\end{addendum}


\clearpage

\begin{table}\centering
  \caption{Magnetic parameters of CrI$_3$, Cr(I,Br)$_3$ and Cr(I,Cl)$_3$ 
  monolayers. Note that (i) the $J=(J_{xx}+J_{yy}+J_{zz})/3$ coefficient listed below is different from the $J'=(J_{\alpha}+J_{\beta})/2$ parameter of Eq. (1), with the former being more reasonable when comparing strength with $D$; and (ii) $S = 3/2$ is used when extracting the magnetic parameters. The energy unit is in meV.}
  \renewcommand\arraystretch{0.8}
  \begin{tabular}{>{\hfil}p{70pt}<{\hfil}>{\hfil}p{80pt}<{\hfil}>{\hfil}p{80pt}<{\hfil}>{\hfil}p{80pt}<{\hfil}>{\hfil}p{80pt}<{\hfil}}
  \hline
  \hline
     Mag. Para.  & CrI$_3$    &   Cr(I,Br)$_3$    &  Cr(I,Cl)$_3$ \\
  \hline
    $J_1$ & -2.218  & -1.800 & -0.983 \\
    $K_1$ & 0.847   & 0.505  & 0.422  \\
    $D_1$ &   0     & 0.270  & 0.191  \\
    $|D_1/J_1|$ & 0 & 0.150  & 0.194  \\
  \hline
    $J_2$ & -0.638   &-0.673  &-0.754 \\
    $K_2$ & 0.078    &0.075   &0.094  \\
    $D_2$ & 0.135    &0.114   &0.138  \\
    $|D_2/J_2|$ & 0.212 &0.169 & 0.183 \\
  \hline
    $A_{zz}$ & -0.262 & -0.124 & -0.029 \\

  \hline
  \hline
  \end{tabular}
\end{table}

\clearpage

\begin{figure}
\centering
  \includegraphics[width=16cm]{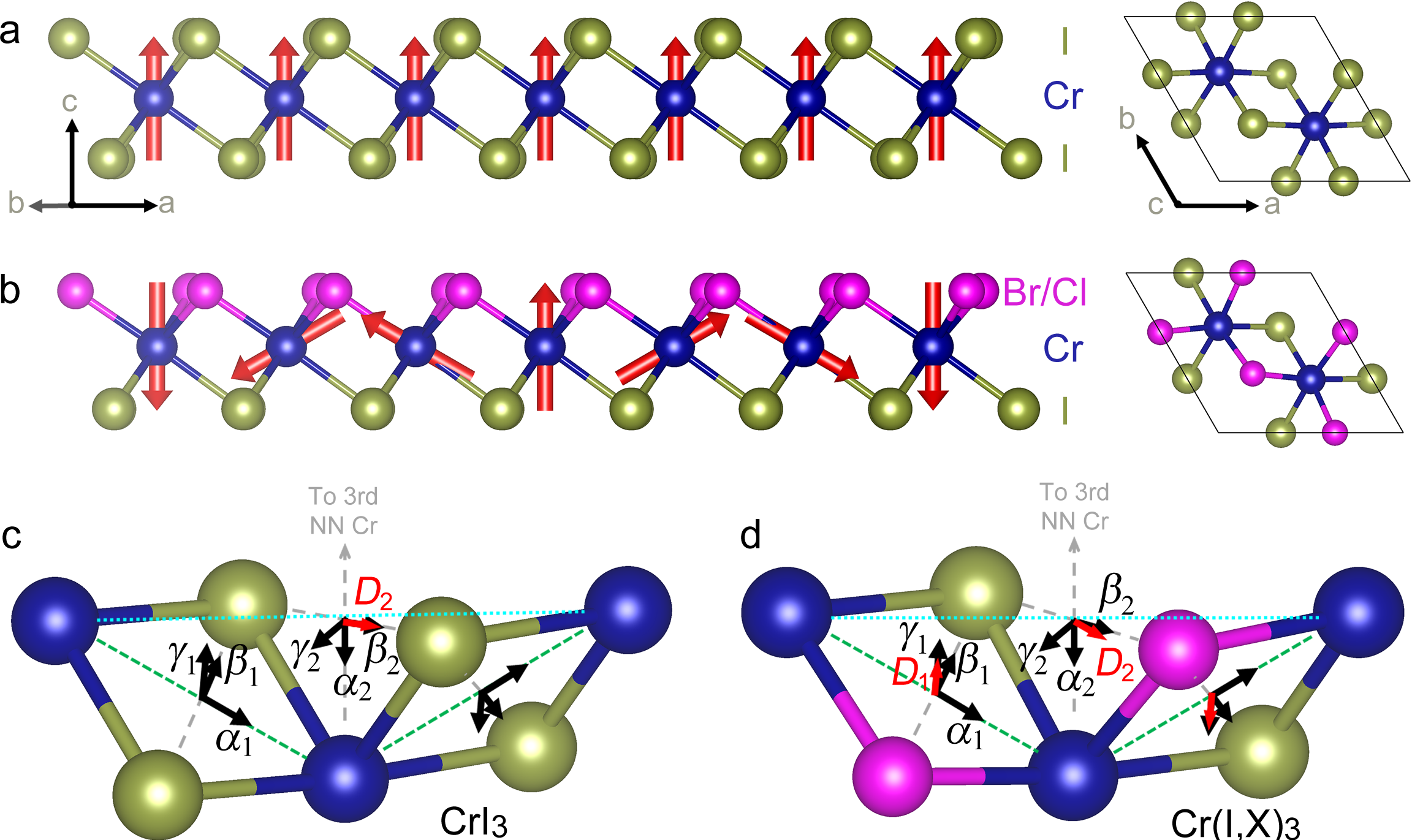}%
  \caption{{\bf Schematics of crystal structure and magnetic interaction vectors of monolayers of CrI$_3$ and Cr(I,X)$_3$.} (a) displays the side and top views of CrI$_3$, with the red arrows schematizing the FM state; (b) shows the side and top views of Cr(I,X)$_3$ (X=Br,Cl), with the red arrows schematizing the helix spins configurations; (c) and (d) illustrate the eigenvectors \{$\alpha,\beta,\gamma$\} of Kitaev interactions and the directions of DMI vectors of CrI$_3$ and Cr(I,X)$_3$, respectively.  }
\end{figure}

\clearpage

\begin{figure}
\centering
  \includegraphics[width=16cm]{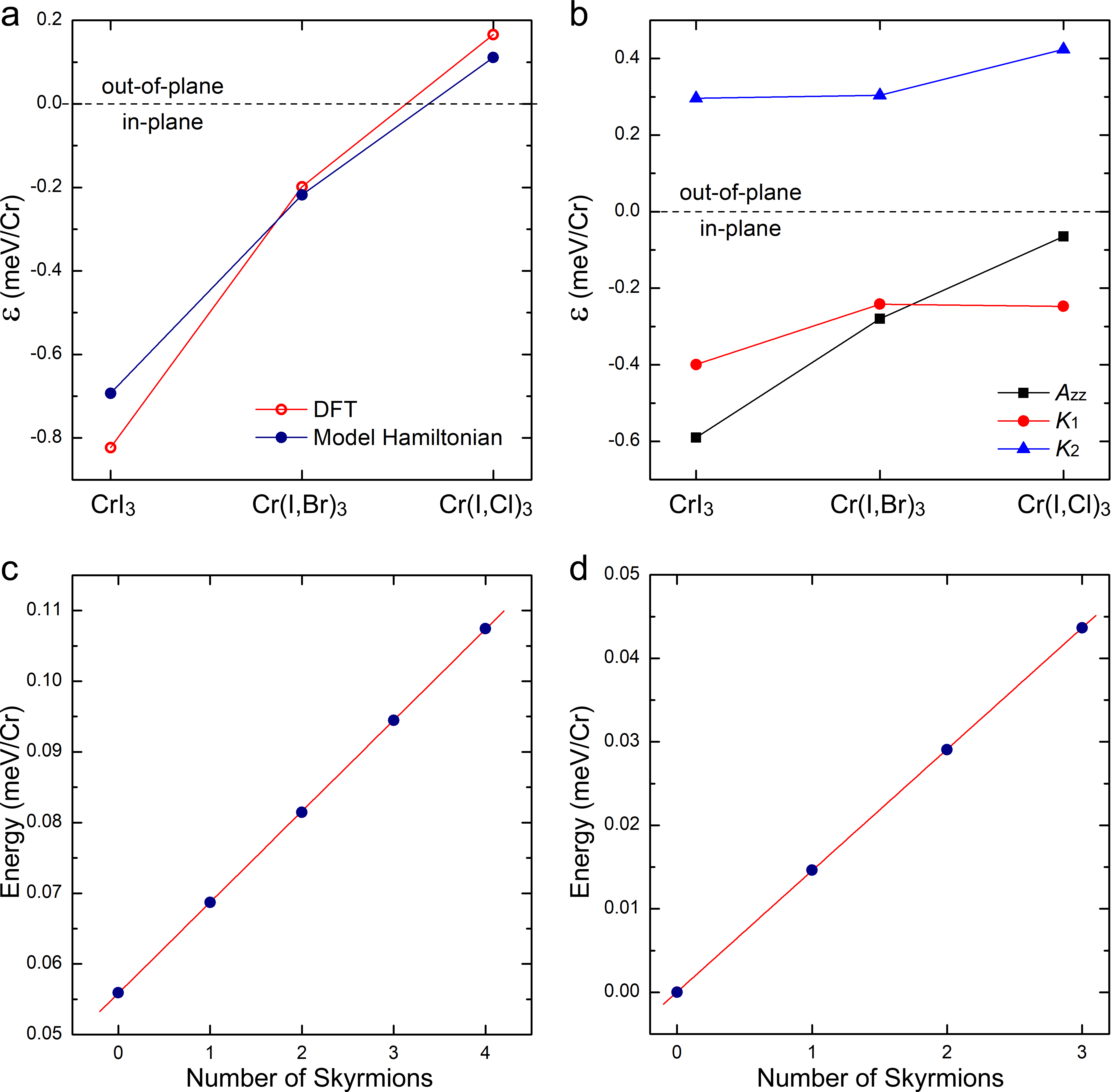}%
  \caption{{\bf Energies of Cr(I,X)$_3$ and CrI$_3$ monolayers.} (a) and (b) show the total and decomposed magnetic anisotropy energy (MAE) of different systems, respectively; (c) and (d) display the evolution of energies as a function of skyrmion numbers for Cr(I,Br)$_3$ and Cr(I,Cl)$_3$, respectively.}
\end{figure}

\clearpage

\begin{figure}
\centering
  \includegraphics[width=16cm]{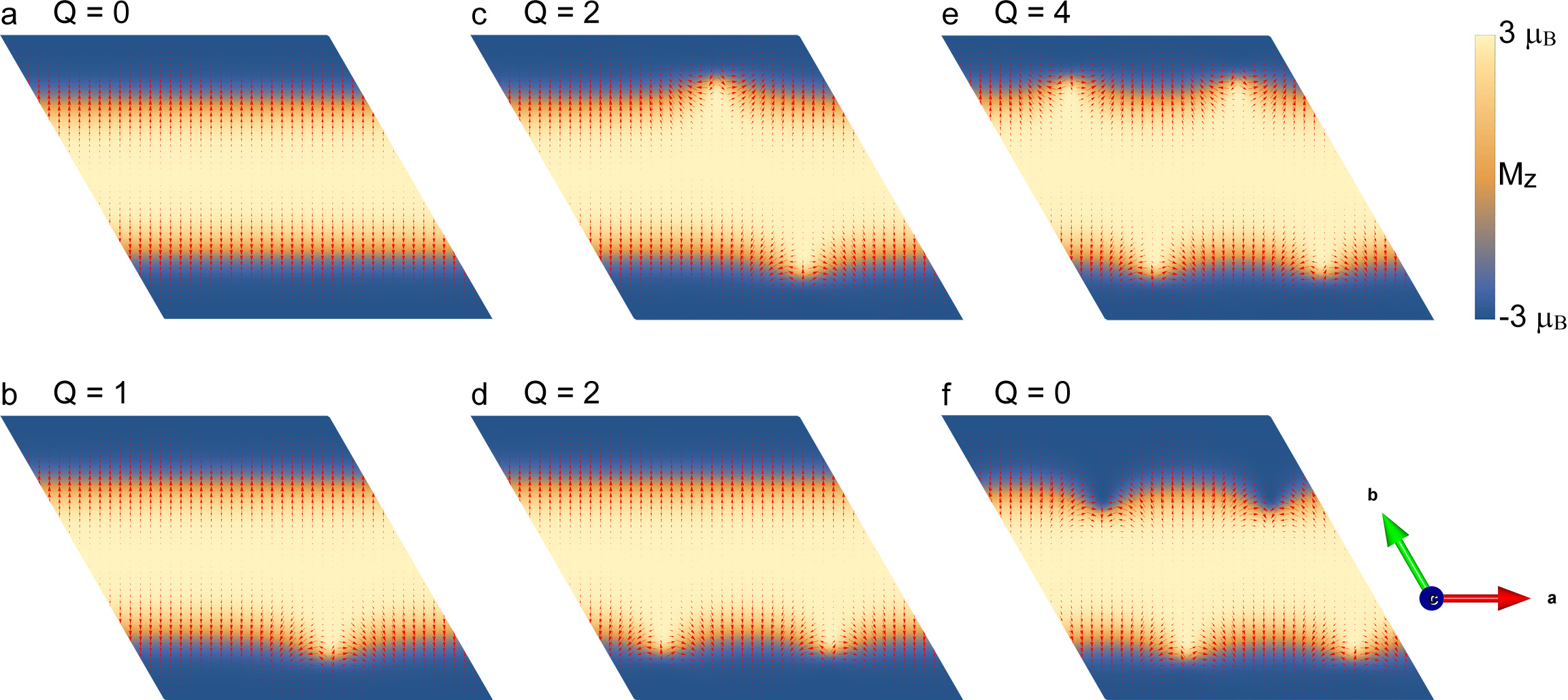}%
  \caption{{\bf Magnetic structures and topological charges of Cr(I,Br)$_3$.} (a) shows the out-of-plane cycloidal structure; (b) illustrates one skyrmion at one domain wall; (c) and (d) display two energetically degenerate structures both with $Q=2$. (e) and (f) each hosts four skyrmions and are also energetically degenerate but with different $Q$. In this Figure,
  the color code applies to the {\it out-of-plane} component of the magnetic dipoles (``spin-in'' and ``spin-out'', for negative (dark) and positive (bright) z-components, respectively) while the arrows characterize the {\it in-plane} components of the magnetic dipoles.}
\end{figure}

\clearpage

\begin{figure}
\centering
  \includegraphics[width=16cm]{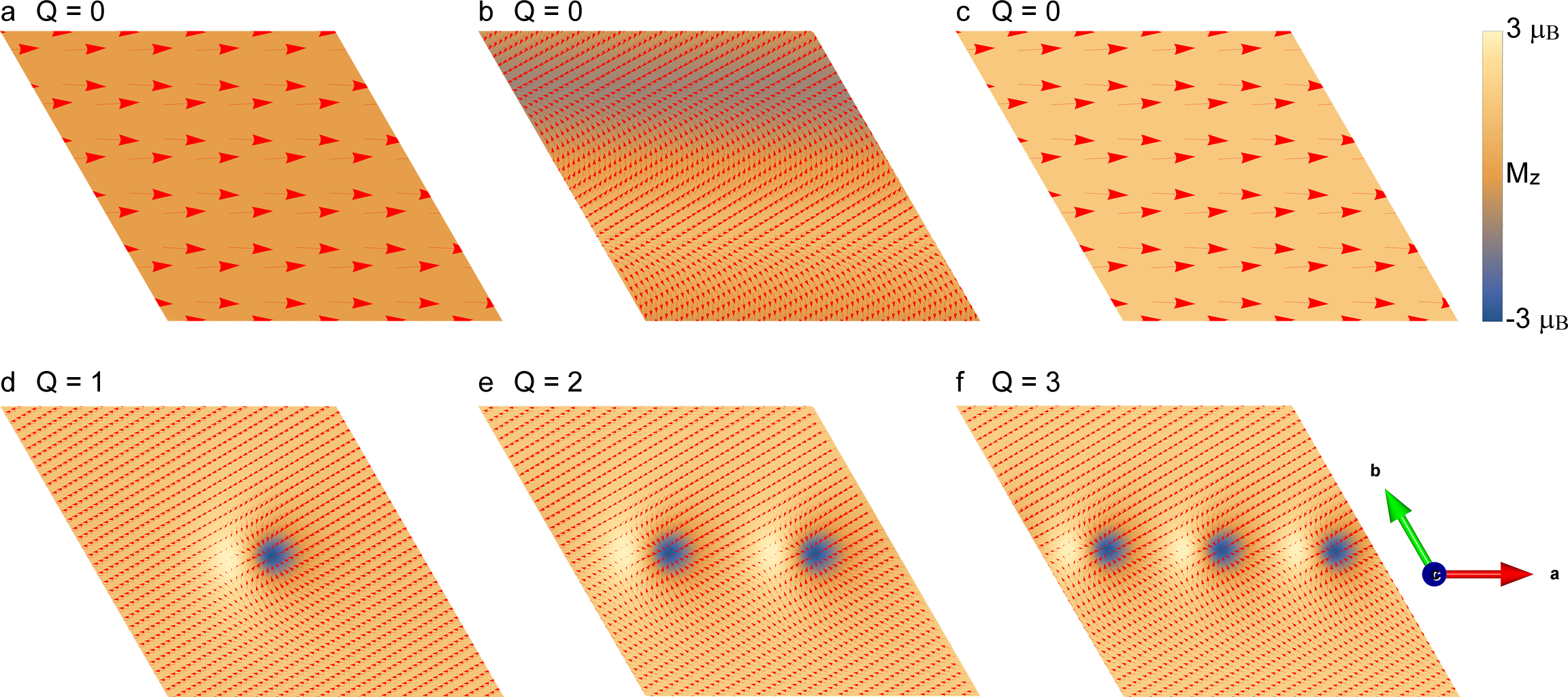}%
  \caption{{\bf Magnetic structures and topological charges of Cr(I,Cl)$_3$.} (a) shows the in-plane zigzag canted FM state; (b) illustrates the in-plane cycloidal structure; (c)-(f) display the cases with an out-of-plane magnetic field $B$ of 0.8 T. (c) also shows the zigzag canted FM state but having an out-of-plane component due to the applied field; (d)-(f) display one, two and three skyrmions, respectively. In this Figure,  the color code applies to the {\it out-of-plane} component of the magnetic dipoles (``spin-in'' and ``spin-out'', for negative (dark) and positive (bright) z-components, respectively) while the arrows characterize the {\it in-plane} components of the magnetic dipoles.}
\end{figure}


\end{document}